\documentclass[letterpaper, 10 pt, conference]{ieeeconf}  % Comment this line out if you need a4paper

\IEEEoverridecommandlockouts  
\usepackage[utf8]{inputenc}
\usepackage[T1]{fontenc}
\usepackage[dvipsnames]{xcolor}
\usepackage{booktabs}
\usepackage{multirow}
\usepackage{graphicx}
\usepackage{subcaption}
\usepackage{amsmath}
\usepackage{amssymb}
\usepackage{mathtools}
\usepackage{cite}
\usepackage[colorlinks=true,allcolors=black]{hyperref}
\usepackage{bm}
\usepackage{listings}           % Source Code Listings. Usage: \begin{lstlisting}...\end{lstlisting}
\usepackage{algorithm,algpseudocode}    % Create pseudo algorithm boxes

\usepackage{mdframed}                   % Breakable and coloured environments
\usepackage[breakable]{tcolorbox}       % Add fancy colored boxes

\tcbuselibrary{skins}
\newtcolorbox{boxy}{drop shadow southeast,enhanced,breakable,colback=white,colframe=black,boxrule=0.5pt,arc=1mm,drop shadow southeast,enhanced}

\usepackage{float}

\usepackage{tikz}
\usepackage{pgfplots}
\pgfplotsset{compat=newest}

\usetikzlibrary{shapes,arrows,decorations.markings}
\usetikzlibrary{arrows.meta}
\usepgfplotslibrary{patchplots}
\usetikzlibrary{positioning}

\tikzstyle{block} = [draw, rectangle, minimum height=2.5em, minimum width=3.5em]
\tikzstyle{cblock} = [draw, rectangle, minimum height=3em, minimum width=3.5em, fill=blue!20,]
\tikzstyle{sum} = [draw, circle, node distance=1cm]
\tikzstyle{circ} = [draw, circle, node distance=2cm]
\tikzstyle{input} = [coordinate]
\tikzstyle{output} = [coordinate] % file with packages and options
\usepackage{import}
\usepackage{xifthen}
\usepackage{pdfpages}
\usepackage{transparent}

\makeatletter
\newcommand{\pushright}[1]{\ifmeasuring@#1\else\omit\hfill$\displaystyle#1$\fi\ignorespaces}
\newcommand{\pushleft}[1]{\ifmeasuring@#1\else\omit$\displaystyle#1$\hfill\fi\ignorespaces}
\makeatother

\newcommand\scalemath[2]{\scalebox{#1}{\mbox{\ensuremath{\displaystyle #2}}}}

\newcommand{\spacing}{~,~~~}

\newcommand{\normal}[2]{\mathcal{N}(#1,#2)}

\newcommand{\var}[1]{\mathrm{Var}}
\newcommand{\re}[1]{{\mathop{\mathrm{Re}}\!}}
\newcommand{\im}[1]{{\mathop{\mathrm{Im}}\!}}

\newcommand{\R}{\mathbb{R}}

\newcommand{\I}{\mathrm{I}}

\newcommand{\itime}{\nu}

\newcommand{\revn}[1] {\textcolor{.}{#1}}

\title{Batch Model Predictive Control for Selective Laser Melting}

\author{Riccardo Zuliani, Efe C. Balta, Alisa Rupenyan, John Lygeros 
	\thanks{
		R. Zuliani, E. C. Balta, 
		A. Rupenyan and J. Lygeros are with the Automatic Control Laboratory, ETH Zurich, 8092 Zurich, Switzerland. A. Rupenyan is also with Inspire AG, 8092
		Zurich, Switzerland {\tt\small$\{$ebalta,ralisa,lygeros$\}$@control.ee.ethz.ch} {\tt\small rzuliani@student.ethz.ch}.
		Research supported by NCCR Automation, a National Centre of
 		Competence in Research, funded by the Swiss National Science
 		Foundation (grant number 180545).} 
}

\begin{document}

\maketitle

\begin{abstract}
% Selective laser melting is a promising additive manufacturing technique that has proven to be able to construct highly customizable high-density products with little material waste. A major challenge in selective laser melting is ensuring the quality of produced parts. It is well understood that the microstructure and the mechanical properties of the produced part depend on its thermal history, hence, to ensure optimal quality, a suitable temperature profile must be generated. Iterative learning control has been proposed as a control technique to obtain suitable laser profiles that produce the desired temperature field profiles. We apply a model-based approach, norm-optimal iterative learning control to achieve the desired temperature distribution. We build on the ILC controller by proposing Batch-MPC (B-MPC), control technique based on the combination of model predictive control and iterative learning control. The selective laser melting dynamics is approximated with a reduced-order control-oriented linear model to ensure reasonable computational complexity. The proposed approache is able to guarantee convergence to the desired temperature field profile despite model uncertainty and repetitive disturbances. Furthermore, since B-MPC is a feedback control technique using in-layer feedback, it is also robust against non-repetitive disturbances.
Selective laser melting is a promising additive manufacturing technology enabling the fabrication of highly customizable products. 
A major challenge in selective laser melting is ensuring the quality of produced parts, which is influenced greatly by the thermal history of printed layers. 
% It is well understood that the microstructure and the mechanical properties of the produced part depend on its thermal history, hence, to ensure optimal quality, a suitable temperature profile must be generated. 
%Iterative learning control has been proposed as a control technique to obtain suitable inputs that produce the desired temperature field profiles. 
% We apply a model-based approach, norm-optimal iterative learning control to achieve the desired temperature distribution. 
% We build on the ILC controller by proposing 
%In this work we propose a control method to reject disturbances on the process.  
We propose a Batch-Model Predictive Control technique based on the combination of model predictive control and iterative learning control. This approach succeeds in rejecting both repetitive and non-repetitive disturbances and thus achieves improved tracking performance and process quality. In a simulation study, the selective laser melting dynamics is approximated with a reduced-order control-oriented linear model to ensure reasonable computational complexity. The proposed approach provides convergence to the desired temperature field profile despite model uncertainty and disturbances.
% Furthermore, since B-MPC is a feedback control technique using in-layer feedback, it is also robust against non-repetitive disturbances.
\end{abstract}

\section{Introduction}

Selective laser melting (SLM) is a promising additive manufacturing technique used to construct parts with highly complex, customizable geometries \cite{W13-1}. In SLM, a part is constructed by sequentially depositing a layer of metallic powder on the bed of a building chamber, which is then melted using a high intensity laser to form a solid part. 
This process is repeated at each layer to form a 3D part \cite{W13-2}. 
Ensuring and maintaining the quality and the repeatability of produced parts through closed-loop control is a major challenge in additive manufacturing, in particular in SLM \cite{renken2019process}. 
The microstructure and the mechanical properties of a part produced by SLM depend on its thermal history. 
However, these properties are not directly measurable in-situ. 
The mechanical properties are often correlated with the temperature distribution on the surface, or with geometrical properties of the melt pool \cite{W8-6}. Furthermore, the layer-to-layer nature of the process poses additional challenges, as defects in the microstructure remain sealed and in most cases cannot be resolved after the corresponding layer is completed \cite{di2021optimizing}. 
Thus, to ensure good quality parts, a suitable temperature profile must be tracked at each layer, which
can be controlled through a suitable laser power profile \cite{W13-5,liao2021layer}. 

Iterative learning control (ILC) has been proposed as a method to control the temperature profile over successive layers by layer-to-layer adjustments of the laser power \cite{W1-7,W14-21}.
ILC aims at improving the performance of a system that operates repetitively by learning from previous iterations, and minimizing an error from a defined control objective. 
One of the advantages of ILC is that it does not require extensive knowledge of the system dynamics \cite{W1-2}. 
ILC is thus suitable for controlling the additive manufacturing processes, while only utilizing reduced-order models of their dynamics \cite{ingyu2017multi,wang2018application,W1-7}. 
ILC relies on an open loop application of an input sequence during an iteration, which is updated only in between successive iterations. 
Consequently, it is able to successfully compensate for static model uncertainties and for disturbance signals that repeat identically in each iteration (or layer in the SLM).
The major drawback of existing ILC techniques applied to SLM is their inability to reject non-repetitive disturbances.
To solve this problem and account for non-repetitive disturbances, \emph{model predictive control} (MPC) can be used in combination with iterative learning control to adjust the control input during an iteration by utilizing run-time measurement feedback \cite{W4-5,W8-4,rosolia2017learning}.

Batch model predictive control (B-MPC) is a receding horizon control approach proposed initially for chemical process control~\cite{W4-4,nagy2003robust}. 
In B-MPC input updates are carried out at each time step by minimizing a quadratic next-iteration cost, which applies a penalty to both the error (or, more precisely, the prediction of what the error is going to be) and the input update signal.
A model-based prediction of the error within an iteration due to the online control updates is used to formulate a finite horizon MPC problem at each time instant for optimal control updates, where input constraints may also be enforced.
The online nature of B-MPC provides robustness to non-repetitive disturbances via the feedback updates during each iteration, \revn{whereas the estimation scheme keeps track of the learned repetitive disturbances by warm starting the error prediction at the beginning of each new iteration. The estimator therefore represents the iterative part of the B-MPC method.}

In this paper, we adapt and apply the B-MPC method of~\cite{W4-4} to control in the temperature distribution in an SLM process, in the presence of both repetitive and non-repetitive disturbances as our main contribution.
The proposed method is validated in simulation and its performance is compared against (i) a proportional controller that uses the same estimation scheme and (ii) a conventional model predictive controller that does not have the estimation updates between iterations to adapt for repetitive disturbances.

The paper is organized as follows: Section \ref{section:II} introduces the control problem, the control-oriented model is presented in Section \ref{section:III}, Section \ref{section:IV} presents the control approach, and Section \ref{section:V} demonstrates the tracking performance of the proposed B-MPC approach. %in comparison with a proportional controller.

\section{Problem description}
\label{section:II}

\begin{figure}
\centering
%\resizebox{5cm}{!}{\incfig[0.7]{drawing}}
%\incfig[0.7]{drawing}
\includegraphics[width=0.65\columnwidth]{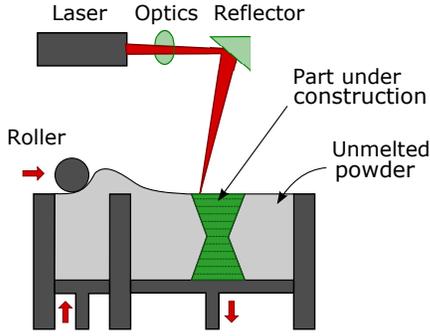}
\caption{Concept of SLM process}
\label{fig:slm}
\vspace{-0.5cm}
\end{figure}

A conceptual illustration of a typical SLM process is shown in Fig.~\ref{fig:slm}. 
A metallic powder is melted by a laser at each layer and then left to solidify to form a solid 3D object.
In this work, we assume that the laser path is fixed and depends only on the geometry of the part in construction. 
Additionally, following~\cite{W1-7}, we consider a simplified thermal dynamics model of the process in which a layer directly connected to the substrate at each iteration. 
With this approximation, the in-layer process dynamics are identical for each layer, the dimensionality of the model are fixed for each iteration $k$, and we have identical initial conditions for each iteration representing a new layer. 
In other words, the temperature distribution of a previous layer does not effect the subsequent layer in this simplified model.

Our goal is to track a desired temperature field profile $T_d(\tau)$, where $\tau\in[0,\tau_f]$ with $\tau_f$ as the layer print time, by adjusting the laser power intensity $u(\tau)\in\mathbb{R}$.
The desired temperature field $T_d(\tau)$ is a time varying vector of temperatures for the whole layer.
For each layer $k$, we consider the temperature tracking control problem
\begin{equation} \label{eq:problem}
    \begin{split}
        \underset{u(\tau): [0, \tau_f]\to \mathcal{U}}{\mathrm{min.}}\quad  &
        \int_{0}^{\tau_f} \|\mathbb{E}[ T(\tau) - T_d(\tau) ]\|_{Q(\tau)}^2 d\tau,\\
        \text{s.t.} \quad & \dot{T}(\tau) = f(T(\tau),u(\tau),d(\tau)), \\
        \quad & T(0) = T_0,
    \end{split}
\end{equation}
where $\mathcal{U}$ is a convex constraint set for the input, $\mathbb{E}[\cdot]$ denotes the expectation of its argument, $T(\tau)$ is the temperature state of the layer at time $\tau$, $\textstyle{d(\tau)}$ is a zero mean i.i.d. multivariate Gaussian noise with covariance $\Sigma_D$, and $f(\cdot, \cdot,\cdot )$ is a Lipschitz continuous function representing the thermal dynamics of the layer.
As we use a reduced order model of the process, we include model uncertainty, in addition to process and measurement noise. 
Additionally, we assume that the desired temperature field $T_d(\tau)$ is identical for all layers (i.e., a fixed trajectory as a function of $\tau$ for all layers and do not depend on the layer index $k$), and the initial condition for each layer is identical, i.e., $\textstyle{T(0) = T_0}$, for all $k$. 
Identical initial conditions is a reasonable assumption whenever there is enough time between layers for sufficient cooling, or for cases when multiple identical layers are printed sequentially over a build plate.

\section{Control-oriented Temperature Modeling for SLM}
\label{section:III}

The temperature dynamics of an SLM process are nonlinear and complex, involving phase transitions and temperature dependent material properties \cite{W13-1}. Finite element models, despite their fidelity, are not suited for control purposes due to their computational complexity \cite{W14-3,W14-4}. In \cite{W1-7}, it is shown how the dynamics of the process below melting point can be well approximated by a linear time-varying system. This linear model is better suited for control purposes because of its simplicity and lower computational complexity. According to \cite{W1-7}, the powder layer can be represented as a grid of $\textstyle{N=n_x\times n_y}$ nodes corresponding to a volume $\textstyle{\delta_x\times\delta_y\times\delta_z,}$ plus an additional node representing the substrate, held at a constant temperature $T_{sub}$. The heat transfer between nodes is expressed via a directed incidence matrix $D\in\mathbb{R}^{N\times q}$, where $q$ is the number of links connecting the nodes. 
Here, we assume the deterministic dynamics of $f$ in \eqref{eq:problem} is linear.
Using this formulation, the heat transfer equation, providing the thermal evolution of the temperature of each node $T(\tau)\in\mathbb{R}^N$ in a single layer, can be expressed as
\begin{align}
    \begin{split}
        C_c\dot{T}(\tau)=&-DKD^{\top}T(\tau)+K_{sub}(T_{sub}-T(\tau))+\\ 
        &+B_c(\tau)u(\tau)~.\label{eq:heattransfer}
    \end{split}
\end{align}
Here, $K=kI_{q}$ and $K_{sub}=k_{\text{sub}}I_{N}$ express the conductivity between nodes and between nodes and substrate respectively, $C_c=cI_{N}$ describes the heat capacity of each node, and $I_n$ denotes the $n$ dimensional identity matrix. $B_c(\tau)\in\mathbb{R}^{N}$ describes, at each time $\tau$, the fraction of the laser power $u(\tau)$ absorbed by each node in the grid. Each node receives a different fraction of the laser power and, overall, all the fractions must sum up to the laser transmission efficiency $\textstyle{\eta\in(0,1]}$. Without loss of generality, we can set $\sum_{i=1}^n B_{c,i}(\tau)=1$ for every $\tau$, i.e., take $\textstyle{\eta=1}$ for simplicity. Equation \eqref{eq:heattransfer} is a reasonably good approximation of the SLM process provided that the melting point is not reached and convection can be neglected.

Due to the modeling assumptions, the steady state temperature distribution is the $T_{sub}$.
Thus, we define all temperatures relative to $T_{sub}$ to simplify the notation by absorbing the affine terms. 
By setting $x(\tau)=T(\tau)-T_{sub}$
we can reformulate \eqref{eq:heattransfer} into a continuous-time state space form:
\begin{subequations}
\begin{align*}
\dot{x}(\tau) &= -A_c x(\tau) +B_c(\tau) u(\tau)\\
y(\tau) &=B_c(\tau)^{\top}x(\tau)\,.
\end{align*}
\end{subequations}
The output of the system is chosen as a weighted average of $x$ of the nodes encompassed by the circumference of the laser beam. After discretizing the system using exact discretization with sampling time $t_s$ we obtain the following discrete-time state space system:
\begin{align}
\begin{split}
    x_k(t\!+\!1)&=Ax_k(t)+B(t)u_k(t)\\ 
    y_k(t) &= C(t) x_k(t) \,,
\end{split}
\label{eq:sysshort}
\end{align}
where $k$ refers to the layer, and $t$ is the time index of each iteration. Assuming all laser scans have an identically finite duration $\tau_f$, we have $t\in[0,\itime]$ with $\itime=\tau_f/t_s$ (assumed for simplicity to be a positive integer). 
Thus, we utilize $y_k(t)$ for the minimization in \eqref{eq:problem} written in $T(\tau)$, so that $Q(t) = C(t)^{\top}C(t)$.
The error is defined as $e_k(t) = y_d(t)-y_k(t)$, hence the objective term in \eqref{eq:problem}, where $y_d(t)=C(t)(T_d(t)-T_{sub})$ is the desired output.
With input disturbances and measurement noise present, system \eqref{eq:sysshort} becomes:
\begin{align}
\begin{split}
x_k(t\!+\!1)&=Ax_k(t)+B(t)u_k(t) +B(t)v_k(t)  \\ 
y_k(t) &= C(t) x_k(t) + w_k(t) \,,
\end{split}
\label{eq:sysnoise}
\end{align}
where we assume that $v_k(t)$ and $w_k(t)$ are Gaussian random variables with zero mean and constant finite variance $\textstyle{v_k(t) \sim \mathcal{N}\left(0,V\right)}$, $\textstyle{w_k(t) \sim \mathcal{N}\left(0,W\right)}$, and are uncorrelated to each other (both across iterations and across time). 
We construct a \emph{lifted representation} of the system output as
$\mathbf{y}_k=[y_k(1)~y_k(2)~\dots~y_k(\itime)]^\top$.
The same can be done for the input, the noises, and the error. 
Assuming, without loss of generality, that $x_k(0)=0$ for all $k$ (since any nonzero initial condition can be seen as a disturbance), we have:
\begin{subequations}
\label{eq:liftshort}\begin{align}
\mathbf{y}_k &= G \mathbf{u}_k +G \mathbf{v}_k + \mathbf{w}_k \\
\mathbf{e}_k &=\mathbf{y}_d-\mathbf{y}_k =\mathbf{e}_{k-1}-G \Delta \mathbf{u}_k + \bar{\mathbf{w}}_k \label{eq:errshort} \,,
\end{align}
\end{subequations}   
where $\Delta \mathbf{u}_k = \mathbf{u}_{k}-\mathbf{u}_{k-1}$ is the input update and
\[
G=
\scalemath{0.7}{
\begin{bmatrix}
C(1)B(0)&\dots&0\\
C(2)AB(0)&\dots&0\\
\vdots&\ddots&\vdots\\
C(T)A^{\itime-1}B(0)&\dots&C(\itime)B(\itime\!-\!1)
\end{bmatrix}\,} .
\]
The term $\bar{\mathbf{w}}_k = -G \mathbf{v}_k - \mathbf{w}_k$ represents layer-wise uncorrelated disturbances with $\bar{\mathbf{w}}_k \sim \mathcal{N}(0,G^{\top}VG+W)$, i.e. disturbances that have no correlation to disturbances in previous or next layers.
We split the error dynamics \eqref{eq:errshort} into its noise-free and noisy parts resulting in
\begin{align}\label{eq:stoc1}
\begin{split}
\bar{\mathbf{e}}_k &= \bar{\mathbf{e}}_{k-1}  - G \Delta \mathbf{u}_k \\ 
\mathbf{e}_k &= \bar{\mathbf{e}}_k + \bar{\mathbf{w}}_k \,.    
\end{split}
\end{align}
In the next section, we propose a controller that iteratively solves \eqref{eq:problem} using the formulation given in \eqref{eq:stoc1} by utilizing the Batch-MPC method proposed in \cite{W4-4}.

\section{Proposed Batch MPC for SLM Processes}
\label{section:IV}

Our control approach leverages the repetitive nature of the process to iteratively learn the optimal laser power profile. To account for model uncertainty, a noise term that is correlated across the layers is introduced in \eqref{eq:stoc1}, next a Kalman filter is derived to construct an error estimate, and finally, an optimization problem is formulated for the input update.

\subsection{Dealing with Model uncertainty}
The error dynamics \eqref{eq:stoc1} is derived from a simplified model \eqref{eq:heattransfer}, leading to a mismatch between the complex physical process and its model used for control. This mismatch leads to a systematic error that, unlike the stochastic noise already present in \eqref{eq:liftshort} is correlated across iterations. To account for this error, a noise term, $\bar{\mathbf{v}}_k$, that is correlated across layers is introduced:
\begin{align}\label{eq:stoc6}\begin{split}
\bar{\mathbf{e}}_k &= \bar{\mathbf{e}}_{k-1}  - G \Delta \mathbf{u}_k + \bar{\mathbf{v}}_{k-1} \\ 
\mathbf{e}_k &= \bar{\mathbf{e}}_k + \bar{\mathbf{w}}_k \, .
\end{split}
\end{align}
The statistical behaviour of $\bar{\mathbf{v}}_k$ depends on the accuracy of the model $G$ as well as the magnitude of $\Delta \mathbf{u}_k$, hence, is difficult to quantify. 
As suggested in \cite{W4-4}, $\bar{\mathbf{v}}_k$ can be modeled as an integrated white noise:
\begin{align} \label{eq:vk}
\bar{\mathbf{v}}_k \sim \normal{0}{\bar{V}} \spacing \bar{V} =
\scalemath{0.7}{\begin{bmatrix}
1&1&1&\cdots&1\\ 
1&2&2&\cdots&2\\ 
1&2&3&\cdots&3\\ 
\vdots&\vdots&\vdots&\ddots&\vdots\\ 
1&2&3&\cdots&\itime
\end{bmatrix}} \sigma_{\bar{V}}^2\,.
\end{align}
If $\bar{\mathbf{v}}_k$ is given as in \eqref{eq:vk}, model \eqref{eq:stoc6} does not describe the error dynamics exactly anymore, however, the statistical behaviour of the noise term $\bar{\mathbf{v}}_k$ can be tuned by varying a single parameter $\sigma^2_{\bar{V}}$. 
Guidelines on the choice of $\bar{\mathbf{v}}_k$, which will impact the filter used to perform state estimation, are given in \cite{W4-4}.

With the model uncertainty, the statistical behaviour of $\bar{\mathbf{w}}_k$ also becomes uncertain, hence, an additional tuning parameter $\sigma_{\bar{W}}$ is introduced:
\begin{equation} \bar{\mathbf{w}}_k \sim \mathcal{N}(0,\bar{W}) \:,\:\:\: \bar{W}=G^\top VG+W+I\sigma_{\bar{W}}^2\,. \label{eq:wbar} \end{equation}
Note, that \eqref{eq:wbar} does not use the same approach as in \cite{W4-4}, instead we leveraged the linear model knowledge ($G$) for approximating the covariance of $\bar{\mathbf{w}}_k$.

To perform a state estimation at each time-step $t$, we partition $G$ as:
\[G \triangleq [G(0),\:  G(1),\: \cdots ,\: G(\itime \!-\!1)],\:\: G(t) \in \mathbb{R}^{\itime}\,,\]
let $\mathbf{e}_k(t)\in\R^{\itime}$ be the error sequence for the $k$-th layer assuming that no input update is performed from time $t$ to time $\itime$ (that is, until the end of the horizon), i.e., \eqref{eq:errshort} with
\begin{align*}
    \Delta {u}_k(t) = \cdots =\Delta {u}_k(\itime\!-\!1)=0\,.
\end{align*}
Then, an equivalent representation of system \eqref{eq:stoc6} in the in-layer domain, i.e., during the iteration, is given by the augmented system in \eqref{eq:stoca}:
\begin{equation}\label{eq:stoca}
\begin{split}
    \begin{bmatrix}
    \bar{\mathbf{e}}_k(t\!+\!1) \\ \mathbf{e}_k(t\!+\!1)
    \end{bmatrix}&=
    \begin{bmatrix}
    I&0\\0&I 
    \end{bmatrix}
    \begin{bmatrix}
    \bar{\mathbf{e}}_k(t)\\ \mathbf{e}_k(t)
    \end{bmatrix}-
    \begin{bmatrix}
    G(t)\\G(t)
    \end{bmatrix}\Delta u_k(t) \, ,\\ 
    e_k(t)& = 
    \begin{bmatrix}
    0& H(t)
    \end{bmatrix} 
    \begin{bmatrix}
    \bar{\mathbf{e}}_k(t)\\ \mathbf{e}_k(t) 
    \end{bmatrix}  \, ,
\end{split}
\end{equation}
where we have 
\[H(t) = [~ \underbrace{~0~~\cdots~~0~}_{t-1} ~~ 1 ~~ \underbrace{~0~~\cdots~~0~}_{\itime-t} ~]\, .\]
Notice that the output of \eqref{eq:stoca} is the measured output of the SLM process $e_k(t)$.
The initial conditions of the state of the system \eqref{eq:stoca}, which also convey the layer to layer dynamics of the system, are given in \eqref{eq:init_aug}.
\begin{align}
\label{eq:init_aug}
\begin{bmatrix}
\bar{\mathbf{e}}_k(0) \\ \mathbf{e}_k(0)
\end{bmatrix}\!=\!
\begin{bmatrix}
I&0\\I&0 
\end{bmatrix}
\begin{bmatrix}
\bar{\mathbf{e}}_{k-1}(\itime)\\ \mathbf{e}_{k-1}(\itime)
\end{bmatrix}+
\begin{bmatrix}
I\\I
\end{bmatrix} \mathbf{\bar{v}}_{k-1} +
\begin{bmatrix}
0\\I
\end{bmatrix} \mathbf{\bar{w}}_{k} \, .
\end{align}
Notice that the impact of the noise terms $\bar{\mathbf{v}}_k$ and $\bar{\mathbf{v}}_k$ is accounted for in the transition between successive iterations in a lifted form.
Therefore, the iteration-wise error states $\bar{\mathbf{e}}_k$ and $\mathbf{e}_k$ capture the effect of disturbances.

\subsection{State estimation for Batch Model Predictive Control}
\label{sec:state_est}
The optimal estimation of the state of the system \eqref{eq:stoca} can be achieved with a Kalman filter. The standard estimation procedure involves repetitive application, at each time step, of the prior update \eqref{eq:prior} and the measurement update \eqref{eq:posterior}.
\begin{align}
	&\scalemath{0.9}{
\begin{bmatrix}
\bar{\mathbf{e}}_{k}(t|t\!-\!1) \\ \mathbf{e}_k(t|t\!-\!1) 
\end{bmatrix}} 
\scalemath{0.9}{= 
\begin{bmatrix}
\bar{\mathbf{e}}_k(\small{t\!-\!1|t\!-\!1}) \\ \mathbf{e}_k(t\!-\!1|t\!-\!1) 
\end{bmatrix}\!-\!
\begin{bmatrix}
G(t\!-\!1)\\ G(t\!-\!1)
\end{bmatrix}\!\Delta u_k(t\!-\!1)} \label{eq:prior} \\
&\scalemath{0.9}{
	\begin{bmatrix}
		\bar{\mathbf{e}}_{k}(t|t) \\ \mathbf{e}_k(t|t) 
	\end{bmatrix}}
	\scalemath{0.9}{= 
	\begin{bmatrix}
		\bar{\mathbf{e}}_k(t|t\!-\!1) \\ \mathbf{e}_k(t|t\!-\!1) 
	\end{bmatrix} + K(t)(e_k(t) -H(t)\mathbf{e}_k(t|t\!-\!1))} \label{eq:posterior}
\end{align}
Where $K(t)$ is the steady-state Kalman gain of the form given in \cite[chap.~16]{bien1998iterative}. The initialization at the beginning of each iteration is:
\begin{align}\label{eq:initall}
\begin{bmatrix}
\bar{\mathbf{e}}_{k}(0|0) \\ \mathbf{e}_k(0|0) 
\end{bmatrix} &= 
\begin{bmatrix}
\bar{\mathbf{e}}_{k-1}(\itime|\itime) \\ \bar{\mathbf{e}}_{k-1}(\itime|\itime) 
\end{bmatrix} \, ,
\end{align}
with $\mathbf{\bar{e}}_{-1}=\mathbf{y}_d$. The repetitive nature of the process is exploited in \eqref{eq:initall}. \revn{The estimator implemented here represents the iterative part of the B-MPC method.}
At the beginning of each new iteration, the error sequence $\mathbf{e}_{k-1}(\itime|\itime)$ containing non-repetitive disturbances is discarded and the new error sequence for the next iteration is initialized considering only the noise with iteration-wise correlation, i.e. the repetitive disturbances.
This way, the controller adapts to non-repetitive disturbances in each layer through state estimation while keeping track of the previously learned repetitive disturbances.

The tuning of the Kalman filter should consider the fact that by giving more weight to $\bar{\mathbf{w}}$ in relation to $\bar{\mathbf{v}}$, e.g. decreasing $\sigma^2_{\bar{V}}$, the change from one iteration to the next one becomes smoother, thus making the algorithm less aggressive and more robust to model uncertainty, at the cost of a lower learning rate. Increasing the value of $\sigma_{\bar{W}}$ produces a less oscillatory in-layer behaviour.

Notice that the term $\bar{\mathbf{v}}_k$ is necessary to force the filter to trade off between prediction and measurement. If $\textstyle{\bar{\mathbf{v}}_k=0}$ then the filter would simply rely on the prediction without considering the measured error and this would prevent the B-MPC algorithm from learning from experience.

\subsection{Batch Model Predictive Control (B-MPC) Formulation}

Here, we present the derivation of a \emph{receding horizon MPC controller}, which computes the optimal input by minimizing a quadratic cost over a finite horizon. The strength of the B-MPC control law comes from its ability to integrate both \emph{online disturbance rejection} (using $\mathbf{e}_k(t|t)$) and \emph{iterative learning} to reject the constant components of the disturbances as well as the model uncertainty (using a non-repetitive noise-free version of the error $\bar{\mathbf{e}}_k(t|t)$ in state estimation).
Thus, the state estimation scheme presented in Section~\ref{sec:state_est} is an essential part of the B-MPC and it is a distinguishing feature compared to conventional MPC.

The control algorithm uses the receding horizon principle, i.e. at time $t$ the MPC minimizes the predicted cost in the next $m$ time steps by choosing an optimal input update. The time horizon $m$ of the MPC must be selected to ensure that the computational complexity is low enough for real-time implementation of the algorithm. After computing the input, which is a sequence of $m$ elements, only the first element is applied. At the next time step $t\!+\!1$ the procedure is repeated.

The B-MPC algorithm approximates an optimal solution to \eqref{eq:problem} by solving an optimization problem with quadratic objective and input constraints over a decision horizon. 
\begin{align}
\begin{split}
    \underset{\Delta \mathbf{u}_k^m(t)}{\mathrm{min.}} & \mathbf{e}_k^{\top}(t\!+\!m|t)Q \mathbf{e}_k(t\!+\!m|t)+ \Delta \mathbf{u}_k^{m,\top}R \Delta \mathbf{u}_k^m(t) \\ 
    \mathrm{s.t.} & \quad 
     \mathbf{u}_{min} \leq \mathbf{u}_k^m(t) \leq \mathbf{u}_{max}\\
     & \quad \delta \mathbf{u}_{min} \leq \mathbf{u}_k^m(t) - \mathbf{u}_k^m(t-1)  \leq \delta \mathbf{u}_{max} \, , \\
     & \quad \mathbf{e}_k(t\!+\!m|t) = \mathbf{e}_k(t|t) - G^m(t) \Delta \mathbf{u}_k^m(t) \, , \\
     & \quad \mathbf{u}_k^m(t) = \mathbf{u}_{k-1}^m(t) + \Delta \mathbf{u}_k^m(t), \\
     & \quad \mathbf{u}_k^m(t-1) = [u_k(t-1)~\mathbf{u}_k^{m-1,\top}(t)]^{\top},
 \end{split}
 \label{eq:BMPC_problem}
\end{align}
where, $\mathbf{u}_{k-1}^m(t)$ is the $m$ control inputs applied starting from time $t$ in the previous layer, $u_k(t-1)$ is the most recent input applied in the current layer, $\Delta \mathbf{u}_k^m(t)$ is the control update in the next $m$ time steps ($t,t+1,\dots,t\!+\!m\!-\!1$):
\[\Delta \mathbf{u}_k^m(t) = \begin{bmatrix}
\Delta u_k(t)& \cdots & \Delta u_k(t\!+m\!-1)
\end{bmatrix}^{\top} \, ,\]
$\textstyle{\mathbf{e}_k(t\!+\!m|t)\in\mathbb{R}^{\itime}}$ is the prediction of the output error at time $\textstyle{t+m}$ considering the future control inputs from $t$ to $\textstyle{t\!+\!m\!-\!1}$,
and $\textstyle{G^m(t) = \begin{bmatrix}G(t) & \cdots & G(t\!+\!m\!-\!1)\end{bmatrix}}$. $Q$ and $R$ are tunable positive semi-definite cost matrices. 
The problem \eqref{eq:BMPC_problem} is a quadratic program (QP) that can be solved efficiently.
Let $\textstyle{\bm{u}_{k}^{m, *}(t)}$ denote the optimizer of \eqref{eq:BMPC_problem}.
Then the first optimal input $\Delta u_k^*(t)$ of the optimal solution is used for the input update, i.e., $\Delta u_k(t) =  \Delta u_k^*(t)$, and the procedure is repeated in a receding horizon.
The control horizon $m$ is fixed as long as $t+m \leq \itime$, otherwise we apply a shrinking horizon of $m=\itime-t$. 
The resulting algorithm to implement the B-MPC is given in \autoref{alg:Algo2}.

\begin{figure}
    \centering
    \resizebox{0.85\columnwidth}{!}{
	    \begin{tikzpicture}[auto,semithick, -latex]
	
		    \node [block] (BMPC) {B-MPC};
		    \node [sum, right of = BMPC, node distance=2.2cm] (sum_inp) {};
		    \node at (sum_inp.center){$+$};
		    \node [sum, right of = sum_inp, node distance=1cm] (sum_v) {};
		    \node at (sum_v.center){$+$};
		    \node [block,right of = sum_v, node distance = 1.25cm] (plant) {Plant};
		    \node [sum, right of = plant, node distance=2cm] (sum_w) {};
		    \node at (sum_w.center){$+$};
		
		    \coordinate [below of= sum_v,node distance = 1cm] (v);
		    \coordinate [right of = v,node distance = 0.7cm] (v2);
		    \draw [draw,-latex] (v2) node [right] {$v_k(t)$} -- (v) -- (sum_v);
		
		    \draw [draw,-latex] (BMPC) -- node [pos=0.5] {$\Delta u_k(t)$} (sum_inp);
		
		    \coordinate [above of = sum_w,node distance = 0.75cm] (w);
		
		    \draw [draw,-latex] (w) -- node[above,pos=0] {$w_k(t)$} (sum_w);
		    \draw [draw,-latex] (plant) -- node [pos=0.5] {$y_k(t)$} (sum_w);
		    \draw [draw,-latex] (sum_v) -- (plant);
		    \draw [draw,-] (sum_inp) -- (sum_v);
		
		    \coordinate [below right = 1cm and 0.35cm of sum_inp] (tmp);
		    \coordinate [right of = tmp,node distance = 0cm] (tmp2);
		    \node [block,below of = tmp2, node distance = 1cm] (est) {Estimator};
		
		    \draw [draw,-latex] ([xshift=+0.3cm]sum_inp.east) -- node[left,pos=0.75] {$u_k(t)$} (tmp) -- (tmp2) -- (est.north);
		
		    \node [sum, below of= sum_w,node distance = 2.15cm] (sum_y) {};
		    \node at (sum_y.center){$+$};
		    \draw [draw,-latex] (sum_w) -- (sum_y);
		    \draw [draw,-latex] (sum_y) -- node [above,pos=0.4] {$e_k(t)$} (est);
		
		    \node at ([xshift=+0.3cm]sum_inp.east) [circle,fill,inner sep=1pt](a){};
		    
		    \coordinate [below of= BMPC, node distance = 2.15cm] (tmp3);
		    \draw [draw,-latex] (est.west) -- node [above,pos=0.5] {$\bar{\mathbf{e}}_{k}(t|t)$} (tmp3) -- (BMPC.south);
		    
		    \coordinate [right of= sum_y, node distance = 0.75cm] (yd);
		    \draw [draw,-latex] (yd) -- node [right,pos=0] {$y_d(t)$} (sum_y);
		
		    \node [block,draw=NavyBlue,above of = sum_inp, node distance = 1.5cm] (data) {$\mathbf{u}_{k-1}=\left\{ u_{k-1}(0), \dots, u_{k-1}(N) \right\}$};
		    
		    \node[text width=2.5cm] at (2.55,2.35) {\textcolor{NavyBlue}{Data block}};
		    
		    \draw [draw=NavyBlue,-latex] (data.south) -- node [right,pos=0.45] {$u_{k-1}(t)$} (sum_inp.north);
		\end{tikzpicture}
	}
\caption{\revn{Block diagram of the B-MPC in closed-loop.}}
\label{fig:block}
\vspace{-0.55cm}
\end{figure}
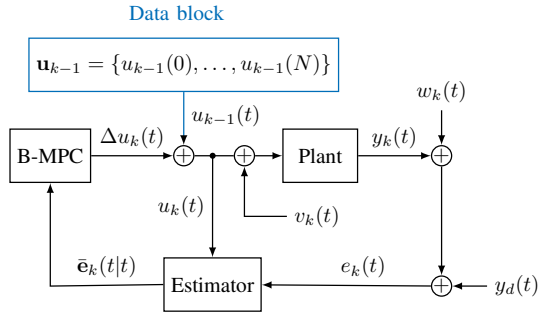

Computation of the optimal input using \eqref{eq:BMPC_problem} accounts for non-repetitive disturbances during an iteration, while the state estimation through the Kalman update keeps track of the learned repetitive disturbances throught the error state. 
A block diagram of the system is presented in \autoref{fig:block}.
\begin{algorithm}
    \caption{Batch MPC}
    \label{alg:Algo2}
    \begin{algorithmic}[1]
    \State \textbf{Init} Compute $G, K(t)$ for $t\in[0,\itime]$
    \State \hspace{0.7cm} Initialize estimator state using $\mathbf{y}_d$
    \For {$k=1:$ \texttt{Maxit}}
        \State Initialize estimator state using \eqref{eq:initall}
        \For {$t=1:\itime$}
            \State Solve QP given in \eqref{eq:BMPC_problem} and compute $\Delta {u}_k(t)$
            \State Apply the input: $u_k(t)=u_{k\!-\!1}(t)+\Delta u_k(t)$
            \State Measure the error: $e_k(t) = y_k(t)-y_d(t)$
            \State Kalman update using \eqref{eq:prior}, \eqref{eq:posterior}
        \EndFor
    \EndFor
    \end{algorithmic}
\end{algorithm}

\section{Case Studies}
\label{section:V}

\begin{figure}
    \centering
    \includegraphics[width=0.85\columnwidth]{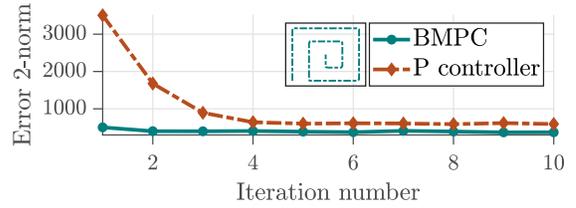}
    \caption{Comparison of the 2-norm errors of the B-MPC and the proportional controller across different iterations.}
    \label{fig:comp}
    \vspace{-0.65cm}
\end{figure}

We demonstrate the B-MPC approach on a simulated SLM process, with the corresponding model and process parameters provided in Table~\ref{table:params}. 
A spiral-in scan path (see \autoref{fig:comp}) has been chosen with a raster spacing of $60\mu m$. %$6\cdot 10^{-5} \, m$. 
The desired output $\textstyle{y_d(t),t\in[0,\itime]}$ has been computed on a system with no model uncertainty by applying a constant power of $20$~W while moving the laser along the diagonal of the grid with a constant velocity of $0.5 \,m/s$. 
The noise covariances $W,V$ have been chosen as $2\%$ of the maximum magnitudes of the desired output and the input respectively.
\begin{table}[ht]
\renewcommand{\arraystretch}{1.1}
\centering
\vspace{-0.55cm}
\caption{Physical parameters used in the simulations}
    \begin{tabular}{@{}lcc@{}}
        \toprule
        \textbf{Parameter} & \textbf{Symbol} & \textbf{Value} \\
        \cmidrule(l{0.15cm}r{0.15cm}){1-3}
        Dimensions of node & $\delta_x,\delta_y$ & $2 \cdot 10^{-5}\, [m]$ \\
        Powder layer thickness & $\delta_z$ & $5 \cdot 10^{-5}\, [m]$ \\
        Conductivity with substrate & $k_{sub}$ & $20\delta_z \, [m]$ \\ 
        Heat capacity of a node & $c$ & $8.5 \cdot 10^{-8} \, [J/K]$ \\
        \bottomrule
    \end{tabular}
\renewcommand{\arraystretch}{1}
\vspace{-0.2cm}
\label{table:params}
\end{table}

The process model \eqref{eq:sysnoise} is used for simulations.
To emulate the effect of the model mismatch, multiplicative uncertainty has been added to the diagonal entries of the heat capacity matrix $C_c$, to the matrix describing the conductivity with the substrate $K_{\text{sub}}$, and to each entry of matrix $B(t)$, $t\in[0,\itime]$ in the simulated model. The random uncertainty is sampled uniformly from the interval $[0,0.3]$ ($[-0.3,0]$ for the uncertainty on $C_c$). For example, the heat capacity matrix is given by $\textstyle{C_{c}=C_{c,\text{nominal}}(I+\text{diag}(\mathbf{r}))}$, where $\mathbf{r}\in\R^N$ is a random vector whose entries are uniformly sampled from the interval $[-0.3,0]$.

\revn{The input magnitude and rate constraints are defined as $u_{\text{max}}=20$~W, $u_{\text{min}}=0$~W, and $\delta u_{\text{max}}=2$~W, $\delta u_{\text{min}}=-2$~W, respectively, with the control horizon $m=20$.}
The Kalman filter has been tuned using the heuristic guidelines given in \cite{W4-4}. 
The resulting parameter values are $\textstyle{\sigma_{\bar{V}}=0.8,\sigma_{\bar{W}}=70}$, and the corresponding sample time is $\textstyle{T_s=10^{-5}\, s}$. \revn{The cost matrices $Q$ and $R$ are tuned heuristically}.

We now compare the tracking performance of B-MPC to the performance achieved with a proportional controller \revn{utilizing the estimation scheme explained in Section~\ref{sec:state_est}}.
The input update commanded by the proportional controller at every time-step is given by:
\begin{align}
	\label{eq:prop_update}
    \Delta u_{k+1}^{c}(t)=k_p H(t) \mathbf{e}(t|t) \, ,
\end{align}
\revn{The proportional gain $k_p$ has been determined heuristically, starting with $\textstyle{k_p=0}$ and increasing until the system reaches instability. 
	Then, the gain has been set to the value that produced the smallest normed error.} 
$\Delta u_{k+1}^{c}(t)$ is then \revn{projected on the constraint set and applied to the system.}

\autoref{fig:comp} shows the corresponding error norm \revn{$||\mathbf{e}_k||$} for the two controllers as a function of the iteration index.
The convergence rate of the B-MPC is better than that of the proportional controller, and the steady state error half of the one obtained with the proportional controller.

\begin{figure}[h]
    \centering
    \vspace{-0.1cm}
    \includegraphics[width=0.75\columnwidth]{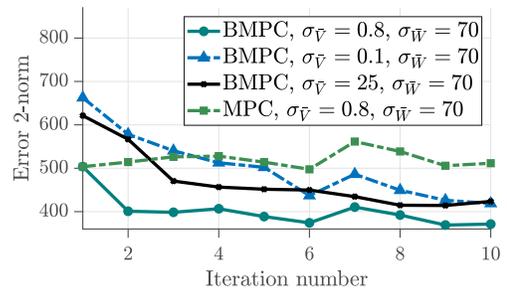}
    \caption{Comparison of the 2-norm errors of the B-MPC for different values of $\sigma_{\bar{V}}$ and with the MPC.}
    \label{fig:compare_noise}
    \vspace{-0.35cm}
\end{figure}

In \autoref{fig:compare_noise} we compare the error norm across different iterations as obtained by the B-MPC with the error norm obtained using a traditional MPC. The MPC update rule is the same as in \autoref{alg:Algo2}, however, at the beginning of each new iteration the input $\mathbf{u}_k$ is reset to the zero vector, and $\mathbf{e}_k$ is reset to $\mathbf{y}_{\text{des}}$. The main difference between B-MPC and MPC is that the latter does not benefit from the estimation of the non-repetitive disturbances, since any information coming from previous iterations is discarded once the iteration ends. Indeed, as can be seen in \autoref{fig:compare_noise}, the error norm obtained using the MPC does not improve over successive iterations. 

In \autoref{fig:compare_noise} we also compare the error norm across different iterations for different choices of the parameter $\sigma_{\bar{V}}$. By decreasing $\sigma_{\bar{V}}$ (e.g. to $0.1$), the KF relies more on the prediction, which is subject to model uncertainty. As a result the convergence rate of the B-MPC decreases and the error norm increases. If $\sigma_{\bar{V}}$ becomes too large (e.g. equal to $25$), the KF relies too much on the noisy measurement, as a result, the B-MPC produces more oscillatory input trajectories, which lead to larger error. 

\begin{figure}[h]
    \centering
    \vspace{-0.15cm}
    \includegraphics[width=0.75\columnwidth]{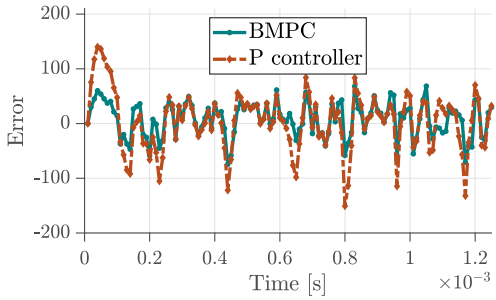}
    \caption{\revn{Comparison of error trajectories in the final iteration.}}
    \label{fig:comp_error_time}
    \vspace{-0.25cm}
\end{figure}

\autoref{fig:comp_error_time} shows the error trajectories corresponding to the two controllers in the final iteration, i.e., ${e}_{10}(t)$ with $t\in[0,\itime]$. 
The proportional controller, which does not consider the constraints, produces larger errors when the laser is in the vicinity of the corners of the spiral path. 
Notice, however, that due to the tightness of the rate constraints, the B-MPC also incurs errors along the corners of the spiral path.

\begin{figure}
    \centering
    \vspace{-0.25cm}
    \includegraphics[width=0.75\columnwidth]{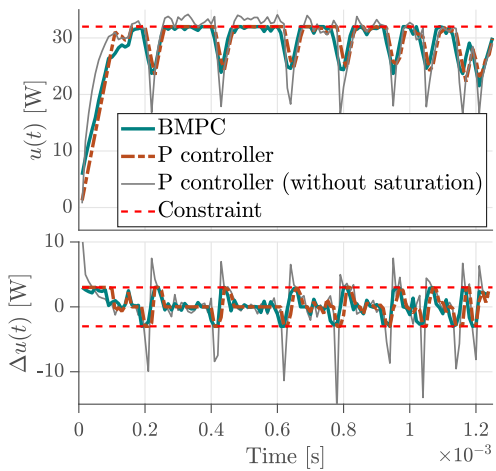}
    \caption{\revn{Comparison of input trajectories in the final iteration.}} 
    \label{fig:comp_input}
%     \vspace{-0.5cm}
\end{figure}

\autoref{fig:comp_input} shows the input applied in the final iteration, i.e., ${u}_{10}(t)$ with $t\in[0,\itime]$. Notice that the commanded input by the proportional controller exceeds the input constraints and is saturated by the system. The lower panel of \autoref{fig:comp_input} shows the magnitude of the updates at each time step, $\Delta {u}_k(t)$. The proportional controller, which does not consider the constraints explicitly, does not always make full use of the entire range of $[\delta \mathbf{u}_{min},\delta \mathbf{u}_{max}]$, due to the saturation.

\section{Conclusion}

In this paper we propose a B-MPC method for closed-loop temperature control of an SLM process.
The B-MPC controller enables both in-layer and between-layers input updates, and achieves good performance even with non-repetitive disturbance. After the end of an iteration, the information about the error and the input of the system is the starting point for the next iteration. The proposed controller is able to obtain high-performance tracking despite repetitive and non-repetitive disturbances and model uncertainty. 
Furthermore, the performance of the B-MPC outperforms the proportional controller in our case study, both in terms of convergence rate and steady-state error.
%To ensure that only the non-stochastic components of the error are used to update the input, we use a Kalman filter to estimate the value of the noise-free error. 

%have applied iterative learning control combined with MPC to achieve correct tracking of a desired temperature profile in a selective laser melting process. 
The SLM process has been modeled using a reduced-order linear time-varying model, which captures the dynamics below melting point. 
%The reduced-order model that has been used monitors only the temperature of the nodes, and it is able to predict the temperature up to the melting point. 
\revn{A possible extension could consider a nonlinear model which takes into account the melt pool dynamics for the process dynamics.}
%	, i.e. the behaviour of the nodes above the melting point, integrating process measurements.}
% A possible extension of such model could take into account the melt pool dynamics, i.e. the behaviour of the nodes above the melting point, integrating process measurements.
Formal convergence analysis of the proposed method with model mismatch and extensions of the control framework with multi-layer models are subject for future work.

%The performance of the B-MPC has proven to be superior to that of a proportional controller, both in terms of rate of convergence and steady-state error.

\bibliographystyle{IEEEtran}
\bibliography{slm_mpc}
	
\end{document}